\documentstyle[aps]{revtex}
\twocolumn

\begin{document}

\title{Liquid Helium and Liquid Neon - Sensitive, Low Background Scintillation 
Media For The Detection Of Low Energy Neutrinos}

\author{D.N. McKinsey and J.M. Doyle}

\address{Department of Physics, Harvard University, Cambridge, MA 
02138, USA}

\date{\today}

\maketitle

\begin{abstract}

The use of liquid helium and neon as scintillators for neutrino 
detection is investigated.  Several unique properties of these 
cryogens make them promising candidates for real-time solar neutrino 
spectroscopy: large ultraviolet scintillation yields from ionizing 
radiation, transparency to their own scintillation light, and low 
levels of radioactive impurities.  When neutrinos scatter from 
electrons in liquid helium or neon, ultraviolet light is emitted.  
The ultraviolet scintillation light can be efficiently converted to 
the visible with wavelength shifting films.  In this way the 
neutrino-electron scattering events can be detected by photomultiplier 
tubes at room temperature.  We conclude that the solar neutrino flux 
from the $\rm p+p\rightarrow e^{+}+d+\nu_{e}$ reaction could be 
characterized and monitored versus time using a 10 ton mass of liquid 
helium or neon as a scintillation target.

\end{abstract}

\section{Introduction}

The observed deficit in solar neutrino flux at the Earth's surface is 
now well established; the neutrino detection rates measured in the 
Homestake\cite{Davis68}, SAGE/GALLEX\cite{SAGE,GALLEX}, and 
Kamiokande/Super-Kamiokande\cite{Kamiokande,Super-K} experiments are 
each significantly less than predicted by the Standard Solar Model 
(SSM), but taken together are also logically incompatible with 
any current solar model.  Resolution of this problem remains 
a tantalizing goal. It is plausible that the correct model explaining
the observed neutrino detection rates involves flavor 
oscillation of massive neutrinos. The several scenarios for 
flavor conversion will most likely be discriminated through 
measurement of the solar neutrino flux, including temporal variations, 
at all energies and for all neutrino species.  Distortions of the 
predicted solar neutrino energy spectra could indicate neutrino flavor 
oscillations, as could daily or seasonal variation of the detected 
neutrino flux.  With these motivations, it is no surprise that 
real-time detection of neutrinos is rapidly becoming more 
sophisticated, with many new detectors either in development or 
recently implemented.

One of the most daunting experimental challenges in neutrino 
observation is the real-time measurement of the full flux of low 
energy neutrinos from the solar reaction $\rm p+p\rightarrow 
e^{+}+d+\nu_{e}$.  This ``pp'' reaction is the most intense source of 
solar neutrinos, and initiates the chain of fusion reactions in the 
sun.  The emitted pp neutrinos range in energy from 0 to 420 keV and 
have a precisely predicted flux of $\rm 5.94\times 10^{10}\, s^{-1} \, 
cm^{-2} $ at the Earth\cite{Bahcall98}.  Despite this high flux, the pp neutrinos have 
proven difficult to characterize in real time; low energy neutrinos 
yield low energy scattering events, and these are difficult to detect 
and discriminate from radioactive backgrounds.  In order to 
characterize and monitor the pp neutrino flux, a detector is needed 
that has a high signal yield for neutrino-induced events, a high rate 
of such events, and a low background rate from intrinsic 
radioactivity.  We are familiar with several approaches to the 
real-time detection of pp neutrinos: bolometric detection of helium 
atoms liberated by rotons from a liquid helium bath 
(HERON)\cite{HERON}, measurement of electron tracks generated in a 
pressurized He (HELLAZ) or $\rm CF_{4}$ (SUPER-MuNu) gas-filled time 
projection chamber\cite{HELLAZ,MUNU}, and the use of a low energy 
neutrino absorbing nuclide that follows absorption with a delayed 
gamma emission(LENS)\cite{LENS}. Here we propose a detector that uses 
liquid helium or neon as a scintillation target. This scheme offers 
the advantages of high scintillation yield, high neutrino detection 
rate, low intrinsic radioactivity, and simplicity.

\section{Experimental Overview}
Detection of neutrinos in our proposed experiment is based on 
neutrino-electron elastic scattering, $\rm \nu_{x} + e^{-} \rightarrow 
\nu_{x} + e^{-}$, where x = ($\rm e$, $\rm \mu$, $\rm \tau $).  For pp 
neutrinos, the scattered electron can range in energy from 0 to 260 
keV. The scattering cross-section for electron neutrinos is about $\rm 
1.2\times10^{-45} \, cm^{2}$ (about 4 times larger than for $\rm \mu$ 
or $\rm \tau$ neutrinos)\cite{Bahcall}.  This small cross-section 
leads to the need for a large detector.  With 10 tons of active 
scintillator ($\rm 3\times 10^{30}$ electrons), a total solar neutrino 
scattering rate of roughly 27 per day will occur with about 18 of 
these from p-p neutrinos (according to the SSM)\cite{Bahcall98}.  This 
mass of liquid helium (neon) fills a 5.1 (2.6) meter diameter sphere.
 
We have diagrammed our proposed experiment in Figure 
\ref{fig:apparatus}.  The design characteristics are similar to those 
used currently in the Borexino experiment\cite{BOREXINO}, with crucial 
differences arising from the choice of scintillator and associated 
cryogenics.  A spherical geometry is chosen for conceptual simplicity 
(a cylindrical volume, for example, could be used instead).

In the center of the experiment is an active region (10 tons) of 
liquid helium or neon. Surrounding the active region is a thin shell 
of transparent material. On the inner surface of this shell is 
evaporated a layer of tetraphenyl butadiene (TPB), a wavelength 
shifting fluor. Around the active (inner) region is a shielding 
(outer) region filled with either liquid neon or liquid helium. If 
neon is used as a shielding medium, it should be about 2 meters 
thick, while if the shielding region is liquid helium, this region 
should be 5 meters thick. These liquids are held in a large 
transparent tank (or 2 separate tanks, see below).

Surrounding the central tank(s), separated 
by vacuum, is another transparent tank filled with 
pure liquid nitrogen.  Outside the cryogens, at room temperature, is a 
large array of low-activity photomultiplier tubes, all facing the 
interior and fitted with light concentrators.  Around the entire 
assembly is a stainless-steel tank, filled with water.

Detection of solar neutrinos is via scintillation originating from 
neutrino-electron scattering that occurs in the active region.  These 
events cause intense emission of extreme ultraviolet light (EUV), 
centered at a wavelength of approximately 80 
nm\cite{Stockton72,Packard70}.  This light is absorbed by the TPB 
waveshifter, causing fluorescence in the blue ($\sim$ 430 nm).  The 
blue light travels through the shield region, through the transparent 
acrylic walls and liquid nitrogen, and is detected by the 
photomultipliers at room temperature.  Detection electronics are 
triggered by multiple photomultiplier coincidence, indicating a 
potential neutrino scattering event.

There are several aspects of this geometry that lead to important 
advantages.  EUV light that originates in the active region will hit the 
TPB film and be converted into blue light, but EUV light that 
originates outside the active region will simply be absorbed and will 
not contribute to the background.  The liquid nitrogen acts both as 
black-body radiation shielding and gamma ray shielding, while the tank of 
deionized water outside the photomultipliers acts as further 
shielding. 

The entire experiment will be located deep underground to reduce 
cosmic ray events.  Muon events will be actively vetoed.  Vetoing 
could be done using a set of photomultipliers to detect Cerenkov light 
in the water tank.  

\section{Signal}

A relatively clear model of scintillations in liquid helium and neon 
can be elucidated from the numerous experimental characterizations of 
charged-particle-induced scintillation in condensed noble 
gases\cite{Stockton72,Packard70,Surko70,Roberts73,Kubota79,Hitachi83,Hab98}. 
 When an energetic charged particle passes through the liquid, 
numerous ion-electron pairs and excited atoms are created.  The ions 
immediately attract surrounding ground state atoms and form ion 
clusters.  When the ion clusters recombine with electrons, excited 
diatomic molecules are created.  Similarly, the excited atoms react 
with surrounding ground state atoms, also forming excited diatomic 
molecules.  Fluorescence in condensed noble gases is observed to be 
almost entirely composed of a wide continuum of EUV light, emitted 
when these excited diatomic molecules decay to the monoatomic ground 
state.  The energy of emission is less than the difference in energies 
between the ground state (two separated atoms) and the first atomic 
excited state for any given noble gas.  The scintillation target is 
thus transparent to its own scintillation light, and a detector based 
on a condensed noble gas can be built to essentially arbitrary size 
without signal loss from reabsorption.

Liquid helium scintillations have been more quantitatively studied 
than neon scintillations.  It has been found that conversion of 
electron kinetic energy into prompt scintillation light is highly 
efficient; about 24\% of the energy of an energetic electron is 
converted into prompt EUV light\cite{Adams98}, corresponding to 15,000 
photons per MeV of electron energy.  Recent work towards detection of 
ultracold neutrons trapped in liquid helium\cite{Doyle94}, has 
resulted in the characterization of efficient wavelength shifting 
fluors that convert EUV light into blue visible light\cite{McK97}.  
This blue light is well matched to the peak sensitivity of available 
photomultiplier tubes.  TPB is the fluor of choice, having a (prompt, 
$\rm <$ 20 ns) photon-to-photon conversion efficiency from the EUV to 
the blue of at least 70\% (and a total conversion efficiency of 
135\%)\cite{McK97,Mattoni}.  The prompt scintillation component from 
the combined liquid helium-waveshifter system has been measured to 
have a 20 ns width, allowing the use of coincidence techniques to 
reduce background\cite{Hab98}.  (In liquid argon and liquid xenon, the 
prompt ultraviolet photon yield has been measured to be even larger; 
Doke \textit{et.  al.} have measured yields of 40,000 and 42,000 
photons/MeV respectively\cite{Doke90}.  This indicates that it is 
likely that neon has a comparable yield.)  Given a scintillation yield 
of 15,000 photons per MeV, a waveshifting efficiency of 70\%, a 
photomultiplier covering fraction of 70\%, and a bialkali photocathode 
quantum efficiency of 20\%, a total photoelectron yield of about 1500 
per MeV could be achieved from the prompt component.  With this 
expected photoelectron yield, the energy of a 100 keV 
neutrino-electron scattering event could be measured with an average 
of 150 photoelectrons, attaining 16\% energy resolution.

Liquid neon can be expected to be a similarly fast and efficient 
scintillation medium, with properties similar to those found in liquid 
helium.  Packard et.  al.  have found that the electron-excited 
emission spectrum of liquid neon peaks at 77 nm\cite{Packard70}. 
Liquid neon should also have an intense afterpulsing component due to 
the extreme ultraviolet radiation of triplet molecules.  In liquid 
helium, the lifetime of this slow component has been measured to be 13 
seconds\cite{McK99}, close to the radiative lifetime of the ground 
state triplet molecule\cite{Chablowski88}. But the theoretically 
predicted lifetime of ground state triplet neon molecules 
\cite{Schneider74} is only 11.9 $\rm \mu s$.  In liquid neon, the 
ground triplet molecular lifetime has been measured to be 2.9 $\rm \mu 
s$\cite{Suemoto79}. Intense afterpulsing following neutrino scattering 
events could be used to positively identify events within the active 
neon, and could also be added into the prompt signal to improve pulse 
height resolution.  However, our detection scheme does not necessarily 
require the use of this afterpulsing signal.

\section{Cryogenics}

We describe here 
the cryogenic and structural requirements for a 
low energy neutrino detector whose active region is a 10-ton 
reservoir of liquid helium or neon. We consider three cases. The 
backgrounds due to construction materials are discussed in section V.

\textbf{Case A: Liquid neon active region, liquid neon shielding 
region.} Here the transparent tank holding the shielding and active 
regions would be constructed of a copper grid and a transparent, low 
radioactivity material, such as quartz or acrylic.  Copper is used to 
give the tank walls high thermal conductivity and structural rigidity, 
while the quartz or acrylic allows scintillation light through to the 
photomultipliers.  Given a total surface area of $\rm \pi (6.6\,m)^{2} 
= 137\,m^{2}$ and a conservatively estimated emissivity\cite{Pobell} of 
1, a total of 270 W is absorbed by the tank walls and routed through a 
copper heat link to a closed-cycle helium gas refrigerator outside the 
shielding.  If the copper grid covers 20\% of the tank surface, has a 
bulk thermal conductivity of 15 W $\rm cm^{-1}$ $\rm K^{-1}$, and this 
copper is 10 cm thick, then the power absorbed from 77 K blackbody 
radiation results in a temperature difference across the tank of no 
more than 2 degrees.  The use of copper to maintain a low thermal 
gradient is necessary because of the narrow temperature window at 
which neon is liquid ($\rm 24.5 - 27.1$ K) and the poor thermal 
conductivity ($\rm \sim 10^{-3}$ W $\rm cm^{-1}$ $\rm K^{-1}$) of 
liquid neon.  The cryogenic constraints on this tank may be relaxed if 
convection in the liquid neon is found to play an appreciable role in 
the flow of heat through its volume.  The active and shielding regions 
are separated by a thin ($\sim$ 0.1 mm) shell of transparent plastic 
or quartz.  This shell simply floats in the neon and is held in place 
by nylon strings connecting the shell to the copper tank.  The shell 
may have small holes in it to allow liquid neon to flow freely between 
the active and shielding regions.

\textbf{Case B: Liquid helium active region, liquid neon shielding 
region.} As in case A, the active and shielding regions are held in a 
copper grid composite tank.  The tank must however be of larger 
diameter (9.1 m instead of 6.6 m) to accomodate the larger active 
region.  Also, the active and shielding regions must be separated by a 
vacuum space because of the different temperatures of the liquid neon 
and liquid helium.  The separation of the active and shielding regions 
must be accomplished with as little material as possible so as to 
minimize radioactive backgrounds.  Appropriate separation may be 
possible using a 1 mm thick Kevlar-acrylic composite shell,
with shielding and active regions held apart using small acrylic 
pegs.   

\textbf{Case C: Liquid helium active region, liquid helium shielding 
region.} Liquid helium is not an effective enough gamma ray 
absorber to protect the active region from copper activity. 
Therefore the tank must be made from a transparent, low radioactivity 
material such as acrylic. The heat load from 77 K is large (1430 W), but 
by cooling the helium through its superfluid transition temperature 
(2.2 K) to achieve high thermal conductivity, the temperature of the 
helium may be made constant throughout its volume. The high thermal 
load on the helium may be handled with a large pumped helium system 
outside the stainless steel tank. As in Case A, the active and 
shielding regions may be separated with a thin sheet of plastic or 
quartz. 

\textbf{General Considerations.} The liquid nitrogen shielding may be 
held in either a copper grid composite or acrylic tank.  The nitrogen 
should be thick enough (1-2 m) to sufficiently absorb gamma rays from 
the photomultipliers and stainless tank.  Acrylic is a low activity, 
transparent, strong material.  At low temperatures, acrylic remains 
strong and tough.  The yield strength of acrylic increases 
significantly as temperature is lowered, while the fracture toughness 
remains roughly constant\cite{Ward}.  Nevertheless, any acrylic 
containers will have to be designed carefully to avoid unnecessary 
thermal and mechanical stresses, as the cryogens are of larger scale 
than is common in low temperature work.

\section{Backgrounds}

Condensed noble gases have an important advantage over organic 
scintillators: they have no $\rm ^{14}C$ contamination.  But among the 
condensed noble gases, only liquid neon and liquid helium can satisfy 
the strictest requirements of low radioactive 
contamination\cite{Seguinot}.  Natural argon is contaminated by the 
two long-lived isotopes $\rm ^{39}Ar$ and $\rm ^{42}Ar$, and natural 
krypton contains $\rm ^{85}Kr$ that precludes its use in low 
background detectors.  Liquid xenon would need to be cleaned of Ar and 
Kr, and double beta decay of $\rm ^{136}Xe$ would have to be 
addressed.  In addition, while liquid xenon has been put to increasing 
use in searches for dark matter, its high price (at least \$1,000,000 
per ton) makes liquid xenon unattractive for use in a large low energy 
neutrino detector.

Helium and neon have no unstable naturally occuring isotopes and 
therefore no inherent radioactive backgrounds.  They do however need 
to be cleaned of dissolved Ar and Kr, as well as possible low-level 
contamination by K, U, and Th, but their low boiling temperatures 
allows for simple and effective solutions to these problems.  
Distillation can effectively remove argon and krypton, and by passing 
the helium or neon through a cold trap, the non-noble radioactive 
contaminants can be frozen out.  In neon one remaining possible 
radioactive contaminant is tritium.  If it is found that commercially 
available neon is contaminated with low levels of tritium, then it can 
be easily removed by chemical means.  Impurities within the helium or 
neon are therefore not expected to be a significant source of 
background.  Helium and neon are also relatively 
inexpensive\cite{neon}.

Because liquid helium and neon are easily cleaned of radioactive 
isotopes, the limiting backgrounds are expected to arise from the 
various construction materials.  Copper (used in cases A and B) has 
been shown to possess low levels of radioactive 
impurities\cite{Avignone86}; an estimate of the activity of copper 
stored underground for a year\cite{HERON} gives .02 events $\rm 
kg^{-1}$ $\rm minute^{-1}$.  Possible impurity levels of other 
necessary materials can be estimated from the results of the 
BOREXINO\cite{Alimonti98} and SNO\cite{SNO} collaborations.  It is 
found\cite{Polycast} that acrylic is commercially available with U and 
Th levels of less than $\rm 10^{-13}$ g/g.  Photomultiplier assemblies 
can be constructed with U and Th levels of $\rm 10^{-8}$ g/g.  Gamma 
rays emitted from the copper, acrylic, photomultipliers, stainless steel tank, 
and heat link will Compton scatter in the nitrogen and shielding 
regions, producing Cerenkov light that can be detected by the 
photomultipliers.  There will be a significant rate of such events; 
for example, the BOREXINO group reports a gamma flux of $\rm 2 \times 
10^{6}\, day^{-1} \, m^{-2}$ from their photomultiplier assembly.  
Fortunately, the light yield from gamma Compton scattering events 
should be relatively small.  Cerenkov light should result in no more 
than 10 photoelectrons per MeV\cite{SNO}, and visible scintillation 
light should contribute even less.  In liquid helium scintillations, 
the visible light output has been measured to be 500 times less 
intense than the extreme ultraviolet output\cite{Stockton72,Dennis69}.  
Furthermore, the visible output is concentrated in wavelengths greater 
than 640 nm, where photocathode responsivities can be chosen to be 
low.  In liquid neon, the visible light emissions are similarly weak, 
with wavelengths that are shifted even further into the 
infrared\cite{Suemoto79}.  As a result, the outer neon region, without 
exposure to an ultraviolet waveshifter, will yield an insignificant 
amount of visible light from gamma scattering events within its 
volume.  However, even with these effects the high rate of gamma 
scattering events in the shielding will produce significant background 
at low photoelectron number.  This will therefore set a low energy 
threshold for neutrino events of roughly 20 keV. This leaves only 10\% 
of solar neutrinos undetected.  With a 2
(5) meter thick liquid neon (helium) shielding region, the rate
of gammas entering 
the active volume should be less than 1/day, compared to the predicted 
27/day solar neutrino counting rate. Also, gamma rays 
that penetrate the shielding region will have relatively high energies 
and are likely to deposit most of their energy in the active region, 
allowing energy cuts to further reduce background. The background levels 
arising from events in the
shielding regions can be independently tested by running 
the experiment without any waveshifter.  

A variety of other effects may help to decrease background counts.  
The three-dimensional photomultiplier arrangement will allow rough 
determination of the event location.  Events in the active volume will 
be more evenly spread over the photomultipliers than events in the 
liquid nitrogen and shielding volume. Also, the light 
concentrators affixed to the photomultiplier tubes will restrict their 
immediate field of vision to the active volume. The 
expected intense ultraviolet afterpulsing from the active liquid neon 
(see section III) could also provide an important test against 
background events.

Radioactive contamination requirements of the materials separating the 
active and shield regions are stringent.  However, very little of 
these materials are necessary.  If clear plastic is used as a divider 
between the active and shielding regions, radioactive background from 
U and Th should be insignificant (given U and Th levels of less than 
$10^{-13}$ g/g.)  However, $\rm ^{14}C$ contamination is a serious 
issue.  In the BOREXINO experiment, $\rm ^{14}C$ levels were 
demonstrated to be less than $1.9 \times 10^{-18}$ $\rm ^{14}C$/C in 
organic scintillator synthesized from petroleum\cite{Alimonti98}.  The 
theoretical estimate for $\rm ^{14}C/C$ in old petroleum is $\sim 5 
\times 10^{-21}$, and the higher measured value is presumed to arise 
during scintillator synthesis or later handling.  A $1.9 \times 
10^{-18}$ $\rm ^{14}C$/C level in a 100 $\rm \mu m$ thick plastic 
divider would result in roughly 80 (30) events per day if helium 
(neon) is used as the active medium.  This would obscure the expected 
27 neutrino events per day.  However, the fact that very little 
material is required ($\sim$ 10 kg of plastic compared to 100 tons of 
organic scintillator used in the BOREXINO experiment) suggests it is 
reasonable to expect that the $\rm ^{14}C$ concentration could be held 
to an acceptable level. In scheme B, a strong, largely transparent 
material is needed to separate the liquid helium and liquid neon 
shielding regions. Because the amount of plastic needed is larger 
than in cases A and C, a lower level of radioactive impurities is 
necessary. 

A second option is to use thin quartz sheet 
as a substrate.  If old silicon is used (older than 50,000 years), 
then $\rm ^{32}Si$ and $\rm ^{14}C$ are not a problem\cite{Martoff}.  
But, of course, $\rm ^{238}U$, $\rm ^{40}K$, $\rm ^{232}Th$, $\rm 
^{3}H$ and $\rm ^{22}Na$ must be shown to contribute less than 1 event 
per day in the energy range of interest.  This should be possible 
because cleanliness levels of less than $10^{-12}$ g/g are routinely 
achieved in pure Si through zone-refining techniques\cite{Cabrera84}.  
By converting this clean Si into silane ($\rm SiH_{4}$) gas, ridding 
the silane gas of radioactive impurities, and then oxidizing, 
sufficiently clean $\rm SiO_{2}$ could be produced.  Again, the fact 
that very little quartz is needed makes this contamination level a 
reasonable requirement.  Contamination requirements on the TPB are not 
so stringent, as only $\rm 0.2 \, mg \, cm^{-2}$ is necessary 
for efficient wavelength shifting\cite{McK97}.

Muons are another potential source of background.  Muons will pass 
through the experiment at a rate of about $\rm 25 \, day^{-1}\, 
m^{-2}$ (at Gran Sasso).  These prompt events can be eliminated 
through active vetoing.  One way to do this is to detect the Cerenkov 
radiation in the ultrapure water tank using a second set of 
photomultipliers\cite{BOREXINO}. In addition, muons that pass through 
the active region will produce extremely bright, easily 
distinguishable scintillation pulses.  

In the neon experiment, neutrons and radioactive species can be 
produced by muons stopping in the active volume.  With only a small 
fraction ($\sim$ .008) of muons stopping\cite{Cassiday73}, and with 
40\% of these stopped muons
absorbed by neon nuclei\cite{Bertin73}, a rate of muon radiogenesis of 
about 0.5 per day follows.  Most of these events result in the 
production of $\rm ^{19}F$, a stable isotope.  Prompt muon coincidence 
rejection and energy cuts will reduce background due to the remaining 
events (e.g. prompt gammas from neutron absorption, decay of long-lived 
nuclei) to negligible levels.  Muons can also lead to the production 
of neutrons in the surrounding rock.  These neutrons, as well as those 
emitted from fission products and ($\rm \alpha$, n) reactions, will be 
moderated and absorbed in the ultrapure water tank, possibly with the 
help of boric acid dissolved in the water\cite{HERON}, and are not 
expected to constitute a significant source of background.

\section{Conclusion}

There are several other experimental programs currently underway to 
develop real time detectors of pp neutrinos.  We believe the method 
described above compares favorably to all of these.  However, making 
exact technical comparisons with HELLAZ, SUPER-MuNu, and LENS is 
beyond the scope of this paper.  Because the HERON experiment also 
uses a liquid cryogen it is possible to make a few simple comparisons.  
The HERON program uses liquid helium as a neutrino scattering medium, 
and bolometers to detect helium atoms liberated by rotons from the 
liquid helium surface\cite{HERON}. The possible 
event rate achievable with HERON is similar to that possible 
using our proposed scintillation technique with helium as the active 
scintillator.  If liquid neon is used, 
however, the event rate is 8 times larger for a given active volume.  
Our design is technically simpler because it requires temperatures of 
only 27 K (2 K) for liquid neon (helium), while HERON requires 30 mK 
superfluid helium to avoid roton scattering.  HERON has the 
requirement (not present in our proposed design) that the helium be 
isotopically pure to avoid $\rm ^{3}He$-roton scattering centers.  The 
added effort and complexity of isotopic purification of 10 tons of 
helium is significant.  A significant technical requirement present in 
our proposed experiment and not in HERON is the need for large, strong 
clear plastic tanks at low temperatures.  Also, unlike HERON, our proposed 
experiment relies almost entirely on high purity shielding materials 
to reduce background, obviating the need for precise event 
reconstruction for background reduction but requiring additional 
materials processing. 

The use of liquid helium or neon as a scintillation medium is a 
promising method for the detection of low energy neutrinos.  First, 
the background level should be very low because of the extreme 
cleanliness possible in the active region.  All other 
materials (with higher levels of contamination) can either be well shielded 
from the active volume or are present in such small amounts that their 
contribution may be made negligible.  Second, the photoelectron output from 
neutrino scattering events should be high because of the intense 
extreme ultraviolet scintillation yield.  Detection with 
standard PMTs is made possible by the availability of efficient 
wavelength shifters.  Third, the rate of detected neutrino scattering 
events will be comparable or larger than those expected in other 
experimental techniques.  Finally, this experiment uses only existing 
technologies; a small ``proof of principle'' apparatus could be 
constructed and tested in relatively little time. 

Along with the calibration and monitoring of the pp neutrino flux, 
this detector will be sensitive to other neutrino sources. For 
example, the relative and absolute intensities of the $\rm ^{7}Be$ and 
pep solar neutrino lines might be measured using this sort of 
detector, yielding a good diagnostic test of what happens to 
neutrinos after they are emitted\cite{Bahcall96}. Whether these line 
intensities could be measured over radioactive background (and other 
neutrino spectra) must be tested by Monte Carlo methods. 

We conclude that liquid helium and neon are intriguing possible 
detectors for solar neutrinos.  An efficient real-time neutrino 
detector based on this technique could be used to calibrate the pp 
neutrino flux from the sun, look for time variation signatures of 
neutrino oscillations, and provide detailed energy information over 
the entire solar neutrino spectrum.

\section{Acknowledgements}

We would like to thank J.N. Bahcall and G.W. Seidel for stimulating 
discussions. This work was supported by National Science Foundation 
Grant No.\ PHY-9424278.

\vspace{1in}

Correspondence Information

Correspondence and requests for materials should be addressed to D.N.M.

\begin{figure}

\caption{Diagram of the proposed experiment: (1) Active region, 
containing 10 tons of ultrapure liquid helium or liquid neon.  (2) 
Sheet of transparent material, coated on its inside surface with TPB 
waveshifter.  In case B there would also be a vacuum region separating 
the active and shielding regions.  (3) Shielding region, filled with 
ultrapure liquid neon or liquid helium.(4) Transparent copper grid 
composite or acrylic tank (5) Ultrapure liquid nitrogen (6) 
Photomultipliers (7) Ultrapure water (8) Stainless steel tank (9) 
Thermal link to refrigerator.  Dimensions assume case A (liquid neon 
active region and liquid neon shielding region.)}

\label{fig:apparatus}
\end{figure}


\begin{references}
\bibitem{neon} Liquid helium can be purchased in small quantities for 
about \$20,000 per ton; liquid neon costs about \$80,000 per ton.
\end{references}
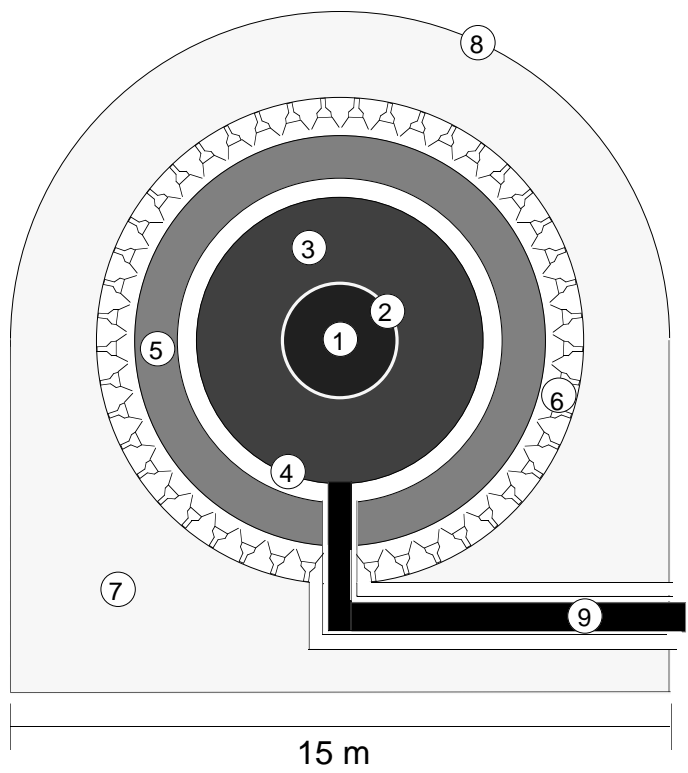
\end{document}